\documentclass[runningheads]{llncs}
\usepackage{multicol}
\usepackage{multirow}
\usepackage[T1]{fontenc}
\usepackage{amsmath}
\usepackage{graphicx}
\usepackage{epstopdf}
 
\usepackage{amssymb} 
\usepackage{array}   
\usepackage{float}
\usepackage{cite}
\begin{document}
\titlerunning{CSAGC-IDS: A Deep Learning Network Intrusion Detection Model}
\title{CSAGC-IDS: A Dual-Module Deep Learning Network Intrusion Detection Model for Complex and Imbalanced Data}
 
%
%
\author{\bf \large Yifan Zeng}
\institute{School of Cyber Science and Engineering, Southeast University, \\Nanjing , China
\\ \email{yifanzeng0615@foxmail.com}}

%
\maketitle  
\vspace{-20pt}
\begin{abstract}
As computer networks proliferate, the gravity of network intrusions has escalated, emphasizing the criticality of network intrusion detection systems for safeguarding security. While deep learning models have exhibited promising results in intrusion detection, they face challenges in managing high-dimensional, complex traffic patterns and imbalanced data categories. This paper presents CSAGC-IDS, a network intrusion detection model based on deep learning techniques. CSAGC-IDS integrates SC-CGAN, a self-attention-enhanced convolutional conditional generative adversarial network that generates high-quality data to mitigate class imbalance. Furthermore, CSAGC-IDS integrates CSCA-CNN, a convolutional neural network enhanced through cost sensitive learning and channel attention mechanism, to extract features from complex traffic data for precise detection. Experiments conducted on the NSL-KDD dataset. CSAGC-IDS achieves an accuracy of 84.55\% and an F1-score of 84.52\% in five-class classification task, and an accuracy of 91.09\% and an F1 score of 92.04\% in binary classification task.Furthermore, this paper provides an interpretability analysis of the proposed model, using SHAP and LIME to explain the decision-making mechanisms of the model.

\end{abstract}
\vspace{-20pt}
\section{Introduce}

\subsection{Background}
With the widespread adoption of network technology, the consequences of cyberattacks have become increasingly severe~\cite{KHRAISAT2019}, and traditional network security techniques~\cite{YUAN2006} are no longer adequate to meet the demands. Network-based Intrusion Detection System can effectively monitor network traffic and detect anomalies~\cite{KHRAISAT2019}. Machine learning, especially deep learning~\cite{LECUN2015,JAVAID2016}, has demonstrated exceptional performance in intrusion detection, but it faces challenges in dealing with imbalanced data and high-dimensional complex data. While deep learning Network-based Intrusion Detection Models (NIDMs) can identify common attacks, their ability to detect rare attacks is insufficient, affecting overall performance~\cite{HE2008,GUPTA2022}. Moreover, deep learning NIDMs still encounter difficulties when handling high-dimensional and complex data~\cite{CUI2023}. High-dimensional traffic data implies a large number of features, complex data patterns, as well as intricate relationships between features, which, along with the increasing complexity of models, pose challenges to the capabilities and structures of deep learning NIDMs. Therefore, further research is needed to enhance the performance of NIDMs in detecting rare attacks under high-dimensional, complex, and imbalanced data conditions.

\subsection{Research Content and Contributions}
In response to the challenge of analyzing high-dimensional, intricate, and imbalanced intrusion traffic data, CSAGC-IDS intrusion detection model is proposed.
\paragraph{SC-CGAN.}
To tackle the issue of data imbalance, the imbalanced data processing algorithm SC-CGAN is proposed. This approach leverages self-attention mechanisms and CNNs to effectively fuse conditional information and capture intricate feature dependencies, ultimately leading to the generation of higher-quality new data. This balanced dataset serves as a valuable resource for subsequent traffic classification tasks. Experimental evaluations have verified that SC-CGAN outperforms other comparative methods.

\paragraph{CSCA-CNN.}For the handling of complexly high-dimensional traffic data, the traffic classification algorithm CSCA-CNN is proposed. This approach integrates channel attention with cost-sensitive learning to extract features and assigns higher costs to minority classes to mitigate imbalanced bias. Experimental results demonstrate that CSCA-CNN surpasses other comparative methods.

\paragraph{CSAGC-IDS.}By integrating SC-CGAN and CSCA-CNN, CSAGC-IDS is constructed. The experimental results indicate that the model surpasses other comparative methods, demonstrating effectiveness and progressiveness in network intrusion detection tasks with high-dimensional, complex, and imbalanced traffic data.

\subsection{Paper Structure}
Section 2 specifically introduces the relevant work in this field. Section 3 details the proposed intrusion detection model, CSAGC-IDS, and its two integral components: SC-CGAN and CSCA-CNN. Section 4 demonstrates the evaluation. It compares the performance of the proposed algorithms and model with existing methods, while also conducting ablation experiments on CSCA-CNN to further analyze its effectiveness. Section 5 concludes the paper with a summary and provides insights into potential directions for future improvements.

\section{Related Work}
\subsection{Deep Learning Intrusion Detection Methods}
Gupta et al. proposed CSE-IDS by combining cost sensitive deep learning and ensemble learning and achieved good performance on imbalanced data~\cite{GUPTA2022}. Li et al. combined multiple CNNs~\cite{LECUN1998} to achieve better accuracy and low complexity~\cite{LI2019multicnn}. Shams et al. proposed CAFE-CNN, which converts traffic data into grayscale images and extracts context aware features~\cite{SHAMS2021CAFEcnn}. Fu et al. combined CNN and bidirectional LSTM to enhance detection performance~\cite{FU2022}. Cui et al. combined CNN and LSTM~\cite{lstmHOCHREITER1997} to form a traffic classifier after extracting features from stacked autoencoder (SAE), fully considering the correlation between data and exhibiting good performance~\cite{CUI2023}.
\subsection{Imbalanced Data Processing Methods}
Synthetic Minority Oversampling (SMOTE)~\cite{CHAWLA2002} can synthesize new minority samples to achieve relative class balance. Jiang et al. used SMOTE to address the data imbalance in network intrusion detection~\cite{JIANG2020}. Ma et al. combined adversarial reinforcement learning with SMOTE for network intrusion detection~\cite{MA2021}. Generative Adversarial Network (GAN) is a generative model proposed by Goodfellow~\cite{GOODFELLOW2014}. Lee et al. oversampled the minority class of network intrusion data using GAN which performed better than SMOTE~\cite{LEE2021}. Douzas et al. used CGAN~\cite{MIRZA2014} to handle imbalanced data, which performed better than other methods~\cite{DOUZAS2017}. Cui et al. used WGAN~\cite{ARJOVSKY2017} combined with GMM for network intrusion detection data balancing, achieving significant performance improvement~\cite{CUI2023}.
\section{Proposed Model for Network Intrusion Detection}

\subsection{CSAGC-IDS Architecture}
CSAGC-IDS consists of two sub module algorithms, SC-CGAN and CSCA-CNN. The former is used for traffic data balancing to reduce imbalance, while the latter classifies and detects traffic. 

\begin{figure}[H]
\centering
\includegraphics[width=1.0\textwidth]{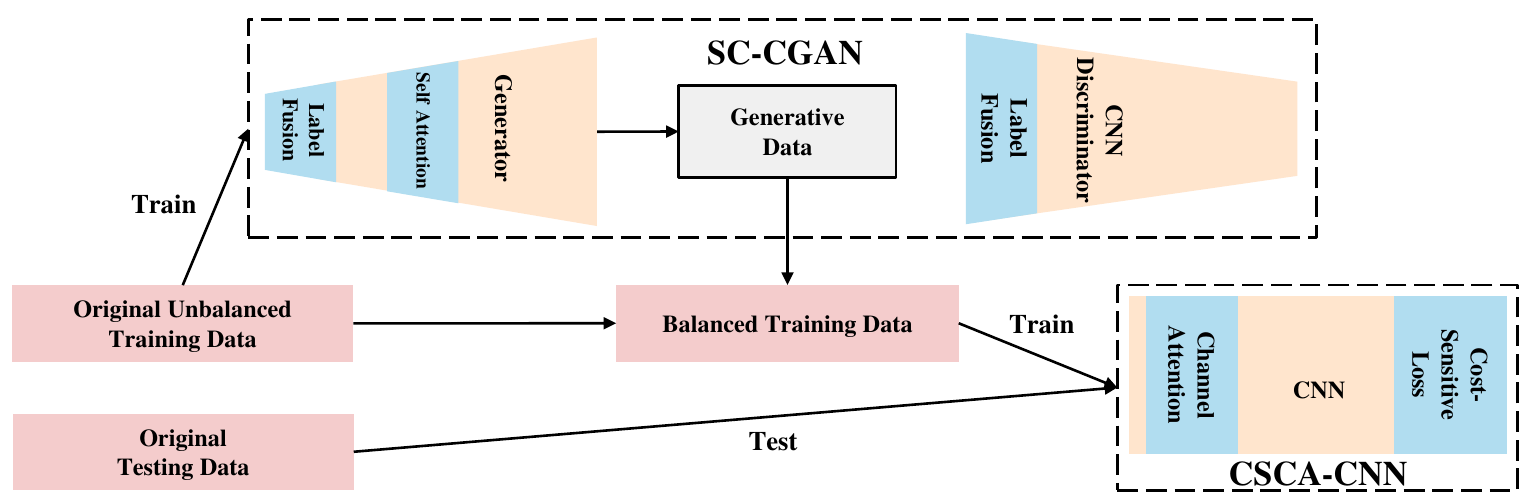}
\caption{CSAGC-IDS architecture} 
\label{fig1}
\end{figure}

The overall architecture is demonstrated in Fig. 1. SC-CGAN uses the original training set to generate new data similar to the original data, and forms a class balanced data with the original training set to train CSCA-CNN. After training, CSCA-CNN was tested to obtain the final detection result. 
The model operation process is demonstrated in Fig. 2.

\begin{figure}
\centering
\includegraphics[width=1.0\textwidth]{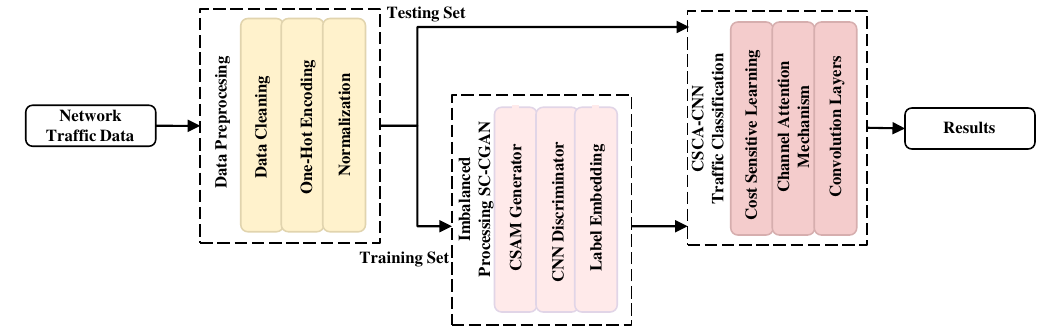}
\caption{CSAGC-IDS process} 
\label{fig1}
\end{figure}
\vspace{-20pt}
Data preprocessing is an important initial step. Numerical processing transforms features into One Hot Encoding that is easily accepted by the model. That only preserves category difference information to avoid misleading the model, allowing the model to better understand the features.Normalization transforms feature values to a certain range, such as [0,1], eliminating the influence of different ranges of feature values. Normalization can make parameter updates more stable and converge faster. Standardization is a kind of normalization:
\begin{equation}  
z = \frac{x - \mu}{\sigma}  
\end{equation}
It converts the original data into a distribution with a mean of 0 and a standard deviation of 1, while retaining the original characteristics of data.

The remaining steps will be introduced below.

\subsection{Imbalanced Data Processing Algorithm SC-CGAN}
SC-CGAN (Self Attention Mechanism Convolution Conditional Generative Adversarial Network) is a generator that integrates a self attention mechanism module~\cite{Transfvaswani2017attention} on the basis of a regular Conditional GAN~\cite{MIRZA2014} , and the discriminator uses CNN~\cite{LECUN1998} to distinguish true or false. The generator, discriminator, and self attention module integrate conditional information into their input, namely the traffic data categorical labels.

SC-CGAN is employed to generate high-quality traffic data, with the objective of balancing the training set, augmenting samples from minority classes, and mitigating model bias stemming from data imbalance. The evaluation results have demonstrated that SC-CGAN exhibits significant advantages over existing methods in the generation of high-quality network traffic data.
The architecture of SC-CGAN is demonstrated in Fig. 3.

\begin{figure}
\centering
\includegraphics[width=1.0\textwidth]{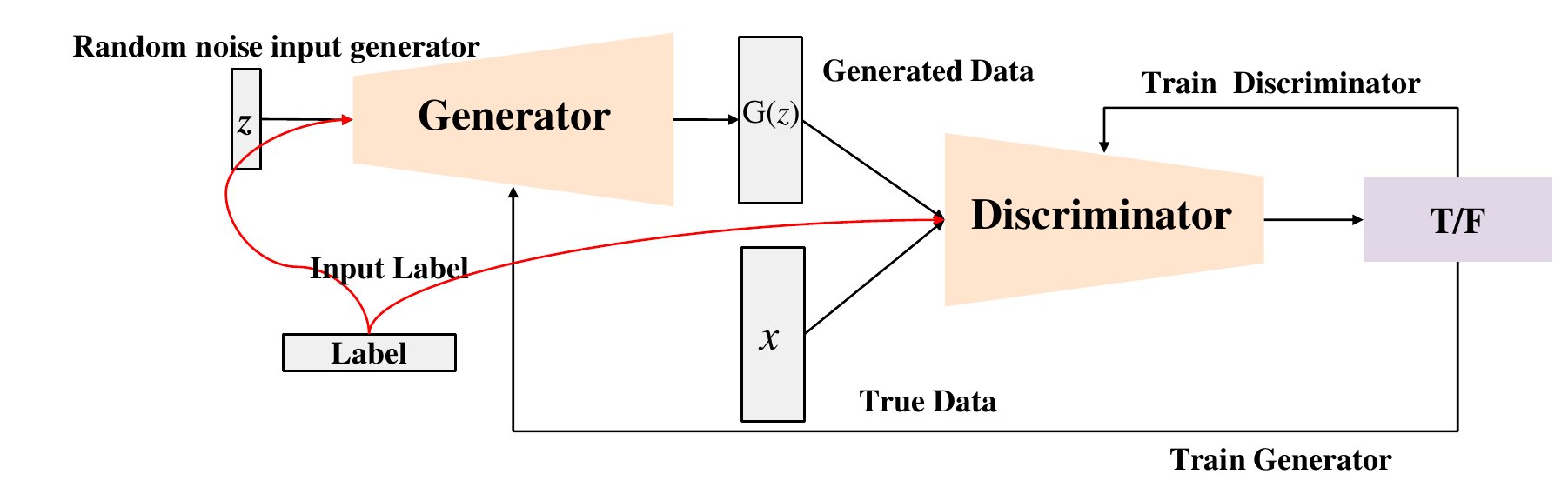}
\caption{SC-CGAN architecture} 
\label{fig3}
\end{figure}

\vspace{-20pt}
\subsubsection{Generative Adversarial Nets with Fusion of Conditional Information.}

While GANs exhibit remarkable generative capabilities, their sole reliance on noise input falls short when dealing with multi-category training data, as it lacks the ability to control the generation of specific categories. To address this limitation, Conditional Generative Adversarial Networks (CGANs)~\cite{MIRZA2014} introduce additional conditional information into both the generator and the discriminator.

SC-CGAN adopts this approach for the generation of traffic data, where the generator integrates conditional information with random noise, and the discriminator combines conditional information with the input sample to be evaluated. Both the generator and discriminator incorporate this conditional information in their respective tasks of generating and discriminating. Specifically, the conditional information in SC-CGAN takes the form of one hot encoded category labels.The loss function of SC-CGAN is defined as follows:

\begin{equation}  
\mathcal{L}_D = -\mathbb{E}_{x, y \sim p_{\text{data}}(x, y)}[\log D(x, y)] -\mathbb{E}_{z \sim p_z(z), y \sim p_{\text{data}}(y)}[\log(1 - D(G(z, y), y))]  
\end{equation}
\begin{equation}  
\mathcal{L}_G = -\mathbb{E}_{z \sim p_z(z), y \sim p_{\text{data}}(y)}[\log D(G(z, y), y)]  
\end{equation}

The generator's loss minimization objective is to produce data that the discriminator deems as authentic (i.e., with an output close to 1), whereas the discriminator aims to minimize its loss by accurately distinguishing between generated data (outputting 0) and real data (outputting 1). Both the generator and discriminator incorporate the conditional information, y, during this process.

Utilizing the category information, the SC-CGAN generator produces samples of specified classes. By generating additional samples from minority classes, it aims to mitigate the imbalance present in the original dataset.

\subsubsection{Conditional Self Attention Mechanism Generator.}

The SC-CGAN generator is integrated with the Conditional Self Attention Mechanism (CSAM). Transformer~\cite{Transfvaswani2017attention} represents a significant milestone in the realm of artificial intelligence, and the SAM serves as its cornerstone. Remarkably, the SAM has not only been implemented in the realm of Natural Language Processing (NLP)~\cite{GPTbrown2020language}, but it has also achieved significant success in the domain of image generation~\cite{selfAttenGANzhang2019self}.
\vspace{-20pt}
\begin{figure}[H]
\centering
\includegraphics[width=1.0\textwidth]{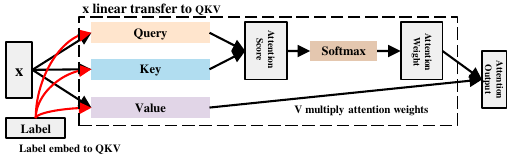}
\caption{CSAM architecture} 
\label{fig4}
\end{figure}
\vspace{-20pt}
The integration of CSAM in SC-CGAN generators is beneficial for generating higher quality traffic data. Fig. 4 demonstrates the CSAM architecture in the SC-CGAN generator, where Query (Q), Key (K), and Value (V) are all obtained through linear transformation of the input. Query retrieves and queries relevant information in the traffic feature sequence. Key is used for similarity matching with Query, while Value is associated with Key. Conditional information is embedded into Query, Key, and Value, and a dot product of Query and Key is computed to determine the similarity between traffic data features. Following a Softmax operation, the attention weight P is obtained and applied to Value for attention-weighted scaling, ultimately yielding the output. The calculation process of the SAM is outlined below:
\begin{equation}
    \text{Attention}(Q, K, V) = \text{softmax}\left(\frac{QK^{\top}}{\sqrt{d_k}}\right)V
\end{equation}

By incorporating a residual connection for this module, can mitigate the issue of vanishing gradients in the model~\cite{Resnethe2016deep}, as illustrated in Fig. 5. Additionally, this residual connection ensures that any raw input information that may be lost during the flow through the CSAM is preserved.
 
\begin{figure}[H]
\centering
\includegraphics[width=0.98\textwidth]{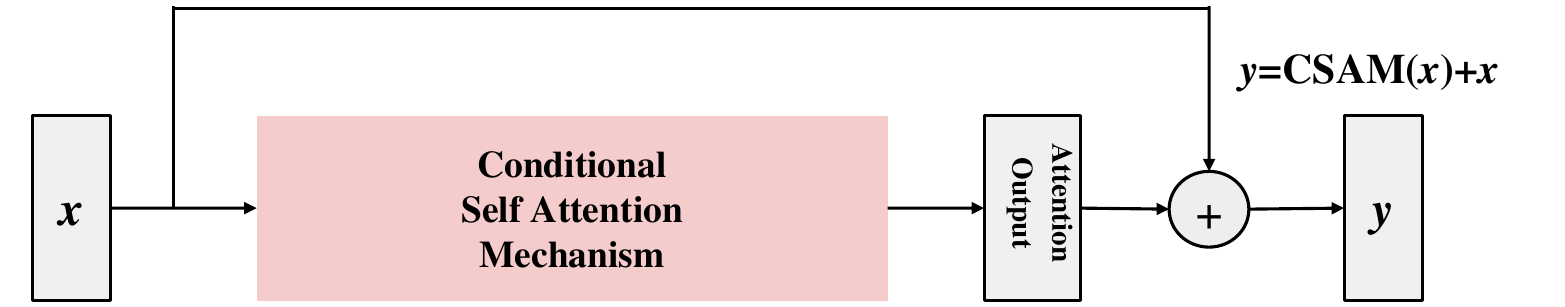}
\caption{Residual connection of CSAM} 
\label{fig4}
\end{figure}

CSAM contributes to the generation of high-quality traffic data in the following significant ways:

\paragraph{Capture Long-Range Dependencies.} CSAM can effectively captures the dependency and correlation relationships among traffic data features, regardless of how far they are in the sequence. Traffic data exhibits numerous dependencies, such as the association between protocol type and port number. When generating traffic data, consider dependency relationships and generate data that matches reality. In complex network scenarios, CSAM is used to adaptively learn dependency patterns.
  
\paragraph{Add Condition Information.} CSAM embeds conditional information into the Q, K, and V. This approach enables more precise control over the generation of specific sample categories.
  
\paragraph{Enhance Model Learning Ability.} Q, K, and V are obtained from learnable parameters. The model introduces more parameters to enhance learning ability. 

\subsection{Traffic Classification algorithm CSCA-CNN}
The CSCA-CNN (Cost Sensitive Channel Attention Mechanism Convolutional Neural Network) framework effectively integrates Cost Sensitive Learning (CSL) ~\cite{GUPTA2022} and Channel Attention Mechanism (CAM) ~\cite{woo2018cbam} within a CNN architecture. CSL addresses the issue of bias towards majority classes, ensuring a more balanced treatment of all classes. On the other hand, CAM enhances the representation of crucial channel features, thereby boosting the overall performance of traffic classification. Fig. 6 illustrates the structure of the CSCA-CNN, showcasing how these two mechanisms are integrated.

\begin{figure}[H]
\centering
\includegraphics[width=1.0\textwidth]{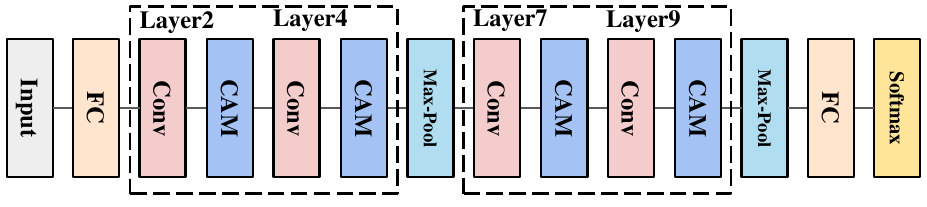}
\caption{CSCA-CNN architecture} 
\label{fig5}
\end{figure}

\subsubsection{Cost Sensitive learning.}

In ordinary classification tasks, it is often assumed that all misclassifications incur an equal cost, but in practical applications, misclassifying instances from different classes can lead to vastly disparate losses. To address this issue, Cost Sensitive Learning (CSL)~\cite{GUPTA2022} has been introduced, which assigns distinct weights to various types of errors and prioritizes the minimization of errors with higher weights during the training process.

CSCA-CNN utilizes CSL to modify the cross-entropy loss function. Specifically, a cost weight matrix is implemented to assign differential weights to the loss functions corresponding to different categories. This approach imposes heavier penalties for misclassifying instances from minority classes, thereby increasing the model's focus on these classes during parameter updates. The cost-sensitive cross-entropy loss function is formulated as follows: 
\begin{equation}
   L = -\sum_{i=1}^{C} w_i \cdot y_i \cdot \log(p_i) 
\end{equation}

\subsubsection{Channel Attention Mechanism.}
CSCA-CNN employs the Channel Attention Mechanism (CAM) feature extraction from the CBAM (Convolutional Block Attention Module) ~\cite{woo2018cbam} framework for traffic classification. This approach aims to enhance the representation of effective and crucial channel features, while minimizing attention to redundant and irrelevant channel features, ultimately improving the overall classification performance.
The specific process of CAM is illustrated in Fig. 7.

\begin{figure}[H]
\centering
\includegraphics[width=1.0\textwidth]{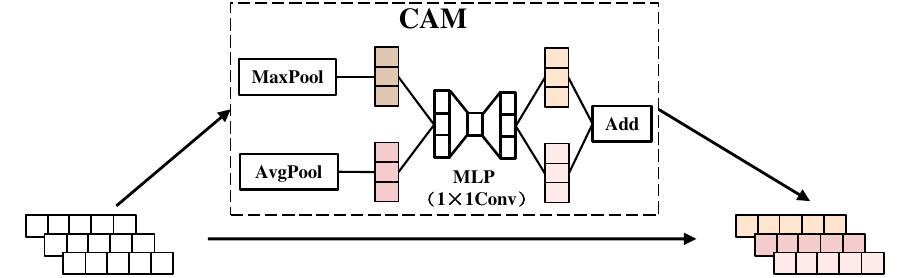}
\caption{CAM architecture} 
\label{fig6}
\end{figure}

\section{Evaluation}
This section conduct extensive experiments to evaluate the proposed algorithms and model in terms of their data generation quality, classification performance, and model complexity. The results obtained validate the advantages of our proposed solution.
\subsection{Experimental Configuration Environment} 
The experiments were conducted on a computing environment with Intel (R) Xeon (R) Gold 6240 CPU @ 2.60GHz. The GPU used was Tesla V100S-PCIE-32GB, and the operating system was Ubuntu 18.04.3 LTS. All code was implemented in Python 3.7.6. The framework was employed PyTorch 1.13.1+cu117.
\subsection{Evaluation Metrics}
Experiments utilize multiple classification performance indicators to provide a comprehensive evaluation, including Accuracy (Acc), Precision (Pre), Recall, and F1-score. Given the significant imbalance in the data, relying solely on Accuracy as a metric is insufficient. Therefore, I also include precision, recall, and F1-score to obtain a more thorough assessment. Notably, the F1-score is particularly valuable as it considers both precision and recall, rendering it a reliable and robust indicator~\cite{CUI2023}.

\begin{equation}  
\text{Accuracy} = \frac{\text{TP} + \text{TN}}{\text{TP} + \text{TN} + \text{FP} + \text{FN}}  
\end{equation} 

In multi-class classification scenarios, it is crucial to consider the global performance across all classes. To achieve this, calculate various class of Pre, Recall, and F1-score separately, and use the ratio of each type of quantity as the weighted average. The F1-score calculation is as follows:
\begin{equation}
    \text{Weighted F1} = \sum_{i=1}^{N} \left( \frac{\text{Number}_i}{\sum_{j=1}^{N} \text{Number}_j} \cdot 2 \cdot \frac{\text{Pre}_i \cdot \text{Recall}_i}{\text{Pre}_i + \text{Recall}_i} \right)
\end{equation}
The indicators for measuring complexity are Params and FLOPs.

\subsection{Dataset}
Experiments utilize the NSL-KDD~\cite{TAVALLAEE2009} benchmark dataset, a widely recognized resource in the field of network intrusion detection. This dataset provides comprehensive and authentic network intrusion traffic data, exhibiting a natural imbalance in data distribution as well as high-dimensional and complex features. These make NSL-KDD an excellent candidate for evaluating the effectiveness and robustness of intrusion detection models. All of the following evaluations were conducted on KDDTest+.
  
\begin{table}[h]    
\centering 
\caption{NSL-KDD Description}\label{tab1}    
\begin{tabular}{|l|l|l|l|}    
\hline    
Class & Description & Quantity & CI Ratio \\    
\hline    
Normal & Normal traffic (no attack) & 77054 & 1 \\    
DoS & Denial-of-Service attack (Overloading to disrupt service) & 53385 & 1.44 \\    
Probe & Probe attack (information gathering) & 14077 & 5.47 \\    
R2L & Remote-to-Local attack (Unauthorized remote access) & 3749 & 20.55 \\    
U2R & User-to-Root attack (attempt to gain superuser privileges) & 252 & 305.77 \\    
\hline    
\end{tabular}    
\end{table}

\subsection{Evaluation of Imbalanced Processing Algorithms}
\subsubsection{Comparative Experiments.}
To measure the quality of traffic data generation for various imbalanced processing algorithms, I evaluate the performance of classifiers trained on data processed by these algorithms. This approach is justified as the performance of the classifier serves as a proxy for the quality of the training data~\cite{DOUZAS2017}. Experiments evaluate the proposed SC-CGAN method and compare it with 8 other imbalance processing algorithms.

\begin{table}[H]
\centering 
\caption{Number of samples generated for each class}\label{tab2}
\begin{tabular}{|l|l|l|l|l|l|}
\hline
Data Source & {Normal} & {DoS} & {Probe} & {R2L} & {U2R} \\
\hline  
Original Data & 67343 & 45927 & 11656 & 995   & 52    \\  
Generated Data &    0  & 21416 & 55687 & 66348 & 67291 \\  
\hline
\end{tabular}
\end{table}

The approach for balancing the experimental data involves generating additional samples for each category and integrating them into the original dataset. This process ensures that the number of samples in each category is equalized. In the case of the KDDTrain+ dataset, since the Normal class originally contained the highest number of samples at 67,343, I generated an equivalent number of samples for each other category to match this figure. The corresponding number of samples generated for each category is presented in the Table 2.

\begin{table}[H]      
\centering      
\caption{SC-CGAN, CVAE, and CSCA-CNN hyperparameters}      
\label{tab:hyperparameters}      
\begin{tabular}{|l|l|l|l|l|}      
\hline      
{Hyperparameters} & {Generator} & {Discriminator} & {CVAE} & {CSCA-CNN} \\ \hline     
Hidden node & 100 & 60 & 60 & 40 \\    
Noise dimension & 123 & - & - & - \\    
Attention dimension & 30 & - & - & - \\      
Activation & LeakyReLU & LeakyReLU & LeakyReLU & LeakyReLU \\     
Initialization & He & Xavier & He & Xaiver \\     
Batch size & 128 & 128 & 128 & 128 \\      
Learning rate & 0.001 & 0.000005 & 0.0001 & 0.01 \\      
Epoch & 30 & 30 & 120 & - \\     
Optimizer & Adam~\cite{kingma2015adam} & Adam & Adam & Adam \\      
Loss function & BCELoss & BCELoss & MSELoss & CSL-CELoss \\       
Convolution kernel size & - & 3 & - & 3 \\      
Dropout & - & 0.3 & - & 0.3 \\     
Maxpool size & - & 2 & - & 2 \\    
Latent dimension & - & - & 32 & - \\    
Number of layers & 8 & 8 & 10 & 12 \\   
CAM squeeze ratio & - & - & - & 8 \\ \hline      
\end{tabular}      
\end{table}

To conduct these comparisons, I implement 5 baseline classifiers and train them using the original imbalanced data, as well as balanced data processed by SC-CGAN and the aforementioned 8 algorithms. Subsequently, I test the classification performance of these trained classifiers to determine the effectiveness of each data balancing method. This comprehensive evaluation allows to gain insights into the quality of data generated by different algorithms and their impact on classifier performance.

The evaluation results, presented in Tables 4 and 5, demonstrate that SC-CGAN exhibits noteworthy advantages across various classification performance indicators. These findings indicate the effectiveness of SC-CGAN in generating balanced training data that significantly enhances the performance of classifiers.
\begin{table}[H]    
\centering    
\caption{Performance of different imbalanced processing algorithms (\%)}  
\label{tab4.8}    
\begin{tabular}{|l|l|l|l|l|l|l|l|l|}    
\hline    
{Algorithm} & \multicolumn{4}{l|}{CNN} & \multicolumn{4}{l|}{Multilayer Perceptron} \\  
\cline{2-9}  
 & Acc & Pre & Recall & F1 & Acc & Pre & Recall & F1 \\    
\hline    
Original Data & 75.44 & 77.44 & 75.44 & 71.74 & 72.14 & 64.70 & 72.14 & 67.45 \\    
ROS & 78.63 & 78.50 & 78.63 & 77.26 & 73.19 & 76.49 & 73.19 & 73.90 \\    
SMOTE~\cite{CHAWLA2002} & 79.40 & 80.58 & 79.40 & 78.27 & 72.37 & 75.93 & 72.37 & 73.64 \\    
Borderline SMOTE~\cite{han2005borderline} & 77.86 & 78.61 & 77.86 & 75.38 & 72.72 & 79.41 & 72.72 & 72.91 \\    
KMeans SMOTE~\cite{douzas2018improving} & 77.45 & 77.67 & 77.45 & 76.01 & 69.54 & 76.76 & 69.54 & 71.74 \\    
SVM SMOTE~\cite{nguyen2009svmsmote} & 79.62 & 80.46 & 79.62 & 77.75 & 75.26 & 77.78 & 75.26 & 75.06 \\    
CVAE~\cite{KINGMA2014,SOHN2015} & 76.75 & 76.56 & 76.75 & 74.31 & 63.40 & 70.12 & 63.40 & 62.68 \\    
CBN-CVAE~\cite{CBNyin2019semantics} & 77.07 & 80.98 & 77.07 & 74.49 & 60.52 & 79.24 & 60.52 & 65.62 \\    
CGAN~\cite{MIRZA2014} & 77.72 & 80.48 & 77.72 & 75.09 & 72.25 & \pmb{82.15} & 72.25 & 75.46 \\    
\pmb{SC-CGAN} & \pmb{80.96} & \pmb{83.04} & \pmb{80.96} & \pmb{78.78} & \pmb{78.28} & 81.74 & \pmb{78.28} & \pmb{78.81} \\  
\hline    
\end{tabular}    
\end{table}

\begin{table}[H]  
\centering  
\caption{Performance of different imbalanced processing algorithms(\%)}  
\label{tab4.8}  
\begin{tabular}{|l|l|l|l|l|l|l|l|l|l|l|l|l|}  
\hline  
Algorithm & \multicolumn{4}{l|}{Decision Tree} & \multicolumn{4}{l|}{Random Forest} & \multicolumn{4}{l|}{K-Nearest Neighbor} \\  
\cline{2-13}
 & Acc & Pre & Recall & F1 & Acc & Pre & Recall & F1 & Acc & Pre & Recall & F1 \\  
\hline  
Original Data & 75.88 & 79.32 & 75.88 & 72.74 & 77.07 & 80.81 & 77.07 & 73.64 & 72.82 & 72.61 & 72.82 & 67.98 \\  
ROS & 77.02 & 79.14 & 77.02 & 73.84 & 76.54 & 81.50 & 76.54 & 73.00 & 74.71 & 78.55 & 74.71 & 71.70 \\  
SMOTE & 76.11 & 78.04 & 76.11 & 73.33 & 75.89 & 80.53 & 75.89 & 72.99 & 75.42 & 78.92 & 75.42 & 73.29 \\  
Borderline SMOTE & 75.89 & 78.73 & 75.89 & 73.59 & 76.63 & 79.91 & 76.63 & 73.59 & 74.94 & 78.41 & 74.94 & 72.08 \\  
KMeans SMOTE & 76.13 & 79.03 & 76.13 & 72.81 & 76.29 & 79.07 & 76.29 & 72.56 & 75.25 & 79.22 & 75.25 & 72.76 \\  
SVM SMOTE & 78.11 & 79.09 & 78.11 & 76.02 & 77.18 & 78.08 & 77.18 & 73.79 & 74.99 & 78.47 & 74.99 & 72.12 \\  
CVAE & 78.45 & 79.10 & 78.45 & 77.18 & 77.36 & 81.71 & 77.36 & 74.07 & 77.87 & 80.07 & 77.87 & 74.85 \\  
CBN-CVAE & 79.40 & 80.69 & 79.40 & 78.19 & 77.78 & 81.42 & 77.78 & 74.95 & 72.57 & 77.10 & 72.57 & 69.52 \\  
CGAN & 79.50 & 80.85 & 79.50 & 76.80 & 77.36 & 81.48 & 77.36 & 75.09 & 78.43 & 81.00 & 78.43 &75.97 \\
\pmb{SC-CGAN} & \pmb{80.19} & \pmb{82.18} & \pmb{80.19} & \pmb{79.01} & \pmb{78.80} & \pmb{82.93} & \pmb{78.80} & \pmb{75.85} & \pmb{79.37} & \pmb{81.29} & \pmb{79.37} & \pmb{76.85} \\
\hline
\end{tabular}
\end{table}

\subsubsection{Dimensionality Reduction for Visualization.}
The process entails diminishing the complexity of high-dimensional traffic data by employing techniques such as t-Distributed Stochastic Neighbor Embedding (t-SNE)~\cite{LAURENS2008} and Stacked Autoencoder (SAE)~\cite{CUI2023}, ultimately projecting the data into a two-dimensional space for intuitive visualization through scatter plots.

The comparative visualization of the original imbalanced data and the SC-CGAN-balanced data, achieved through the utilization of t-SNE (depicted on the left) and SAE (displayed on the right) dimensionality reduction algorithms, is presented in Fig. 8.
\begin{figure}[H]
\centering
\includegraphics[width=1.0\textwidth]{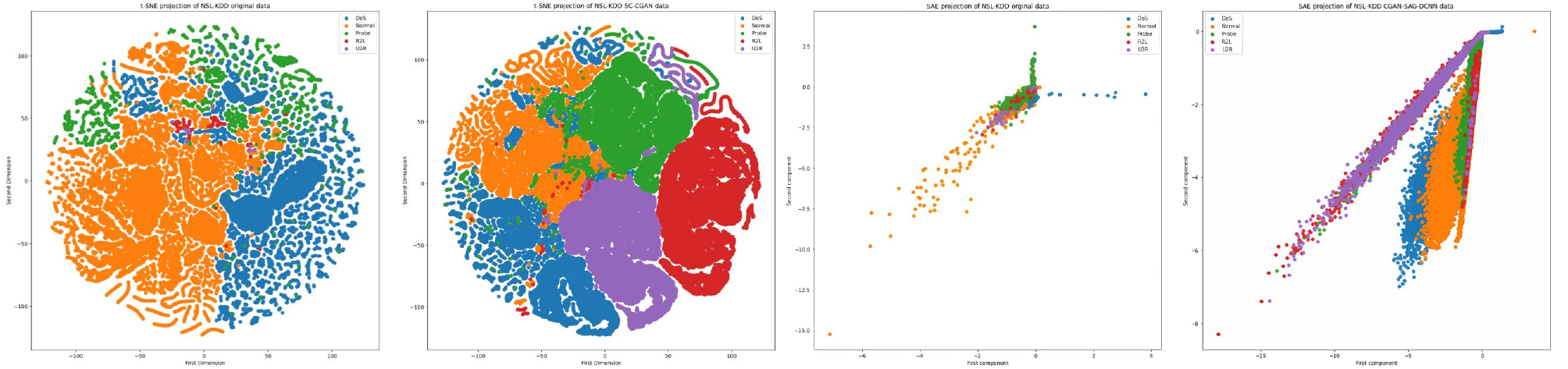}
\caption{NSL-KDD original imbalanced data and SC-CGAN balanced data dimensionality reduction visualization by t-SNE and SAE} 
\label{fig7}
\end{figure}

Observing the results, it is evident that the balanced data generated by SC-CGAN notably augments the presence of samples from rare classes compared to the original data, thus effectively mitigating data imbalance. Consequently, the decision boundaries between various classes become more distinct, favoring the classification task. Furthermore, the data augmentation achieved through SC-CGAN widens the data coverage, potentially enhancing the model's generalization capabilities.
\subsection{Evaluation of Traffic Classification Algorithms}

All classification algorithms are trained on the balanced data generated by SC-CGAN. To assess the effectiveness of the CSL and CAM components, ablation experiments are conducted. Following this, a comparative analysis is performed between the baseline classification algorithm and the proposed CSCA-CNN. Finally, a comparative evaluation of the algorithm complexity of CSCA-CNN is undertaken with respect to the complexity reported in other relevant studies.

\begin{table}[H]      
\centering      
\caption{Ablation experiment results for CSCA-CNN(\%)}    
\label{tab4.8}      
\begin{tabular}{|l|l|l|l|l|l|l|l|}      
\hline      
Algorithm & CSL & CAM & CNN & Acc & Pre & Recall & F1 \\   
\hline    
\pmb{CSCA-CNN} & \checkmark & \checkmark & \checkmark & \pmb{84.55} & \pmb{85.70} & \pmb{84.55} & \pmb{84.52} \\ 
CNN-Only & - & - & \checkmark & 80.96 & 83.04 & 80.96 & 78.78 \\    
w/o CAM & \checkmark & - & \checkmark & 82.66 & 83.37 & 82.66 & 82.30 \\    
w/o CSL & - & \checkmark & \checkmark & 81.72 & 83.41 & 81.72 & 79.60 \\  
\hline      
\end{tabular}      
\end{table}  

\begin{table}[H]      
\centering      
\caption{Comparative experiment results between CSCA-CNN and baseline classifiers(\%)}    
\label{tab4.8}      
\begin{tabular}{|l|l|l|l|l|}      
\hline      
Classifier & Acc & Pre & Recall & F1 \\  
\hline 
Naive Bayes & 53.80 & 48.44 & 53.80 & 44.25 \\  
Logistic Regression & 77.04 & 79.68 & 77.04 & 73.67 \\  
K-Nearest Neighbor & 79.37 & 81.29 & 79.37 & 76.85 \\  
Decision Tree & 80.19 & 82.18 & 80.19 & 79.01 \\  
Random Forest & 78.80 & 82.93 & 78.80 & 75.85 \\  
XGBoost~\cite{CHEN2016} & 78.94 & 81.31 & 78.94 & 76.73 \\  
Multilayer Perceptron & 78.28 & 81.74 & 78.28 & 78.81 \\   
\pmb{CSCA-CNN} & \pmb{84.55} & \pmb{85.70} & \pmb{84.55} & \pmb{84.52} \\   
\hline      
\end{tabular}      
\end{table}
\vspace{-20pt}
\begin{table}[H]      
\centering      
\caption{Comparative experiment results on complexity of CSCA-CNN}    
\label{tab4.8}      
\begin{tabular}{|l|l|l|l|l|l|l|l|}      
\hline 
Indicator & CNN & 1D-CNN & DNN 2 layers & DNN 3 layers & DNN 4 layers & DNN 5 layers & \pmb{CSCA-CNN} \\  
\hline 
Cite & \cite{DING2018} & \cite{LI2023} & \cite{LI2023} & \cite{LI2023} & \cite{LI2023} & \cite{LI2023} & - \\
 Params & 126826 & 90373 & 841221 & 1235717 & 1366789 & 1399557 & \pmb{49469} \\  
FLOPs & - & 6886280 & 1680670 & 2469150 & 2731038 & 2796446 & \pmb{729000} \\ 
\hline      
\end{tabular}      
\end{table}
The ablation study unequivocally demonstrated the effectiveness of the proposed enhancement. In the realm of traffic classification, the CSCA-CNN stands out, exhibiting considerable superiority compared to conventional baseline classifiers. Notably, the CSCA-CNN boasts lower Params and FLOPs, translating to reduced storage requirements and a diminished risk of overfitting. Furthermore, its efficient design ensures it necessitates less computational resources, making it a cost-effective and efficient solution for traffic classification tasks.

\subsection{Evaluation of Intrusion Detection Models} 
In evaluating the performance of the CSAGC-IDS, a comprehensive comparison is conducted with both classical models and the start-of-the-art artificial intelligence models that have been proposed by researchers in recent years. Initially, I analyze the binary classification capabilities of the CSAGC-IDS, distinguishing between normal traffic and attack patterns. Subsequently, I delve deeper into comparing the performance of the five-class classification models, which categorize traffic into five distinct classes: Normal, DoS, Probe, R2L, and U2R. 

\subsubsection{Binary Classification.}
To evaluate the binary classification performance of the CSAGC-IDS model, a comparison has been conducted against various benchmark models, including LR, NB, SVM-rbf, DNN 1 layer, DNN 5 layers~\cite{VINAYAKUMAR2019}, Multi-CNN~\cite{LI2019multicnn}, and DLNID~\cite{FU2022}.
\begin{table}[H]      
\centering      
\caption{Models binary classification performance comparative results(\%)}    
\label{tab4.8}      
\begin{tabular}{|l|l|l|l|l|}      
\hline      
{Model} & {Acc} & {Pre} & {Recall} & {F1} \\   
\hline 
LR~\cite{VINAYAKUMAR2019} & 82.60 & {91.50} & 74.40 & 82.00 \\  
NB~\cite{VINAYAKUMAR2019} & 82.90 & 86.50 & 80.50 & {83.40} \\  
SVM-rbf~\cite{VINAYAKUMAR2019} & 83.70 & 76.90 & \pmb{99.30} & 86.70 \\  
DNN 1 layer~\cite{VINAYAKUMAR2019}  & 80.10 & 69.20 & 96.90 & 80.70 \\  
DNN 5 layers~\cite{VINAYAKUMAR2019} & 78.90 & 68.00 & 96.30 & 79.70 \\  
Multi-CNN~\cite{LI2019multicnn} & 86.95 & 89.56 & 87.25 & 88.41 \\  
DLNID~\cite{FU2022} & 90.73 & 86.38 & 93.17 & 89.65 \\  
\pmb{CSAGC-IDS} & \pmb{91.09} & \pmb{93.68} & 90.45 & \pmb{92.04} \\   
\hline      
\end{tabular}      
\end{table}
\vspace{-20pt}
\subsubsection{Five-Class Classification.}
For Siam-IDS~\cite{BEDI2019}, I-SiamIDS~\cite{BEDI2021}, and LIO-IDS~\cite{GUPTA2021}, where only various classes of Pre, Recall, and F1-score are provided, I utilized weighted approach to calculate overall indicators for comparison.

\begin{table}[H]      
\centering      
\caption{Models five classification performance comparative results(\%)}    
\label{tab4.8}      
\begin{tabular}{|l|l|l|l|l|}      
\hline      
{Model} & {Acc} & {Pre} & {Recall} & {F1} \\   
\hline 
J48~\cite{TAVALLAEE2009} & 81.05 & - & - & - \\  
NBTree~\cite{TAVALLAEE2009} & 82.02 & - & - & - \\  
RandomTree~\cite{TAVALLAEE2009} & 81.59 & - & - & - \\  
SVM~\cite{TAVALLAEE2009} & 69.52 & - & - & - \\  
AlexNet~\cite{JIANG2020} & 77.02 & 78.54 & 77.24 & 77.88 \\  
LeNet-5~\cite{JIANG2020} & 79.91 & 82.95 & 80.01 & 80.45 \\  
BiLSTM~\cite{JIANG2020} & 79.43 & 81.14 & 79.65 & 80.39 \\  
DNN 5 layers~\cite{VINAYAKUMAR2019} & 78.50 & 81.00 & 78.50 & 76.50 \\  
CNN~\cite{DING2018} & 80.13 & - & - & - \\  
Multi-CNN~\cite{LI2019multicnn} & 81.33 & - & - & - \\  
CAFE-CNN~\cite{SHAMS2021CAFEcnn} & 83.34 & 85.35 & 83.44 & 82.60 \\ 
SCAD-RNN~\cite{Singh2019} & 82.61 & - & - & - \\ 
Siam-IDS~\cite{BEDI2019} & - & 77.39 & 77.41 & 75.65 \\  
I-SiamIDS~\cite{BEDI2021} & - & 78.77 & 80.32 & 78.81 \\  
LIO-IDS~\cite{GUPTA2021} & - & 81.13 & 80.80 & 80.77 \\ 
DQN~\cite{SETHI2020} & 81.80 & - & - & - \\  
SSDDQN~\cite{DONG2021} & 79.43 & 82.81 & 79.43 & 76.22 \\  
AE-RL~\cite{CAMINERO2019} & 80.16 & 79.74 & 80.16 & 79.40 \\  
AESMOTE~\cite{MA2021} & 82.09 & 84.11 & 82.09 & 82.43 \\ 
AE-SAC~\cite{LI2023} & 84.15 & 84.27 & 84.15 & 83.97 \\
\pmb{CSAGC-IDS} & \pmb{84.55} & \pmb{85.70} & \pmb{84.55} & \pmb{84.52} \\   
\hline      
\end{tabular}      
\end{table}
\vspace{-20pt}
Based on the results presented in Tables 9 and 10, the CSAGC-IDS demonstrates exceptional performance, highlighting its progressiveness and effectiveness. The CSAGC-IDS excels in learning deep feature representations of data, making it adept at managing complex, high-dimensional, and imbalanced data compared to other deep neural network architectures.

\subsection{Interpretability Analysis}
While CSAGC-IDS exhibits remarkable performance, its deep learning structure and vast parameters hinder interpretability~\cite{ADADI2018}, a crucial aspect in network intrusion detection. Given the high error costs, administrators often require more than just labels to make informed decisions~\cite{WEI2023}.

To enhance interpretability, I utilize LIME~\cite{Ribeiro2016} and SHAP~\cite{Lundberg2017} on the NSL-KDD to dissect CSAGC-IDS to provide with insights into the model's decision-making process, fostering trust and enabling detection of potential errors.

\vspace{-5pt}
\subsubsection{LIME.}
LIME (Local Interpretable Model-agnostic Explanations) trains interpretable models to approximate complex decision boundaries on individual samples, providing localized explanations~\cite{Ribeiro2016}.

For the binary classification task of CSACG-IDS on the NSL-KDD, LIME employs a linear model to mimic the CSACG-IDS. The coefficients of this interpretable model quantify each feature's impact on predictions. I select 4 normal samples and generate 50,000 perturbed samples for analysis.
\vspace{-10pt}
\begin{figure}[H]
\centering
\includegraphics[width=1.0\textwidth]{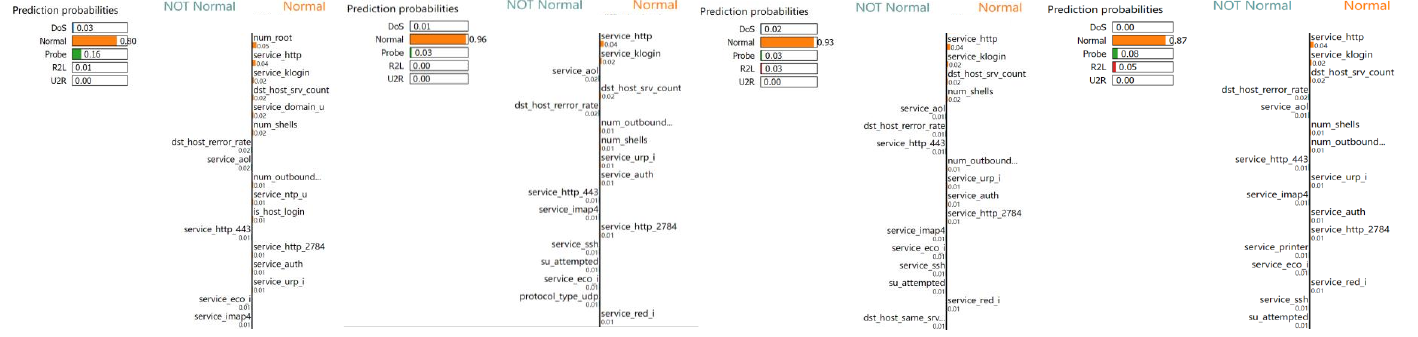}
\caption{4 normal samples interpretability analysis by LIME} 
\label{fig7}
\end{figure}

\vspace{-10pt}
As illustrated in Fig. 9, by integrating the analysis of 4 normal samples, it is evident that features such as {service\_http}, {dst\_host\_srv\_count}, {service\_klogin}, {service\_urp\_i} and {num\_shells} have a positive effect on predicting normal classes in CSACG-IDS. Conversely, features such as {service\_aol}, {service\_eco\_i}, {service\_imap4} and {service\_ssh} have a negative effect. The service\_http represents common HTTP services that appear frequently in the Internet. It may indicate behaviors such as DoS. The dst\_host\_srv\_count reflects the usage of services on the host, which may reveal attempts at port scanning and exploiting vulnerabilities in multiple services. The feature service\_klogin indicates Kerberos login, which may suggest legitimate user authentication behavior. The num\_shells may reveal that malicious users or scripts are attempting to control the system, which could be related to U2R and R2L attacks. The service\_ssh represents the SSH service used for remote management, which may reveal the existence of unauthorized remote access and control (U2R, R2L).

\vspace{-5pt}
\subsubsection{SHAP.}
SHAP (SHapley Additive exPlanations) assesses each feature's contribution, quantifying its average marginal impact on the model's outcome via the Shapley value, offering a quantitative analysis for each feature~\cite{Lundberg2017}.

SHAP offers a binary interpretation for CSAGC-IDS, visualizing the Shapley values of each feature in a force plot (Fig. 10). Here, the Shapley values are depicted as forces, indicating their impact on the results, with red indicating positive contributions and blue representing negative effects. In prediction of the attack sample, features such as {service\_http} have a positive effect on predicting attack class and features such as {flag\_REJ} have a negative effect. 
\vspace{-20pt}
 \begin{figure}[H]
\centering
\includegraphics[width=0.7\textwidth]{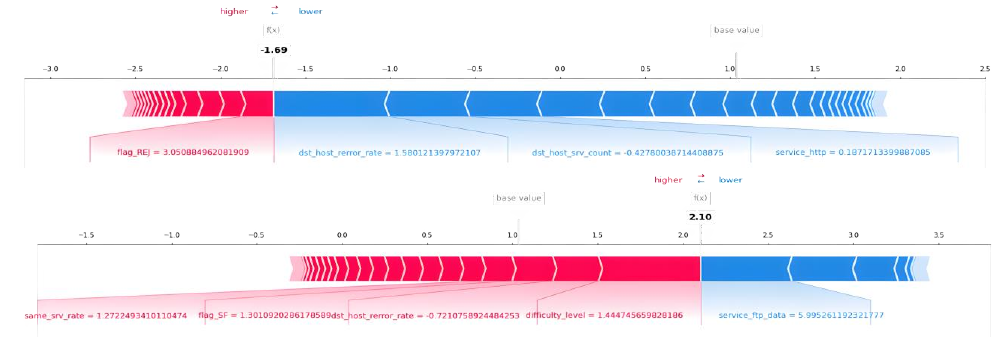}
\caption{Attack and normal samples SHAP force plot} 
\label{fig7}
\end{figure}
\vspace{-20pt}
By taking 100 samples and plotting force plot in horizontal stack. As illustrated in Fig. 11, Sample 56 is classified as an attack sample, based on the values of features including {service\_http}, {dst\_host\_rerror\_rate} and {service\_imap4}. The service\_imap4 may reveal behaviors that exploit the IMAP4 mail service for attack. The dst\_host\_rerror\_rate indicates the rate of connection errors on the destination host, which could suggest that the host is under substantial invalid or malicious requests (DoS).

\vspace{-20pt}
\begin{figure}
\centering
\includegraphics[width=0.7\textwidth]{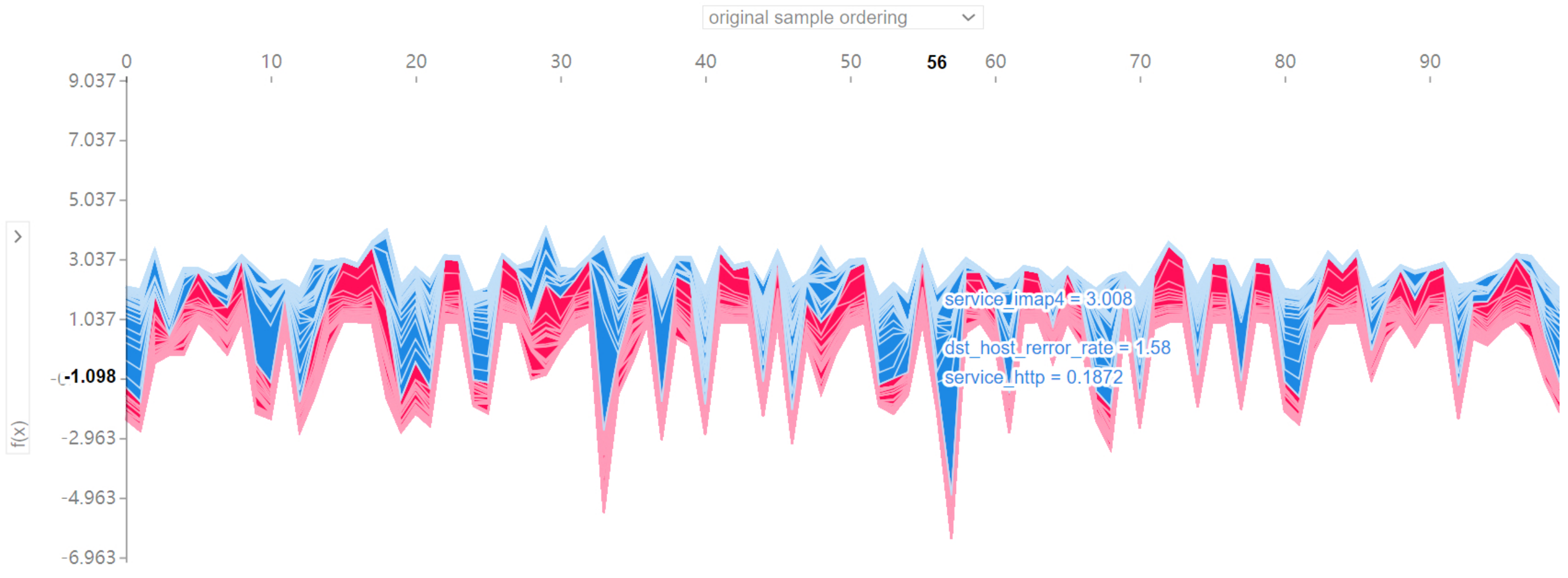}
\caption{100 samples SHAP force plot} 
\label{fig7}
\end{figure}

\vspace{-20pt}
\section{Conclusion}
CSAGC-IDS, a deep learning network intrusion detection model that leverages cost sensitive learning and a mixed attention mechanism to tackle the challenges of high-dimensional, complex, and imbalanced data distributions in network intrusion detection. Experimental results demonstrate its effectiveness in achieving superior performance for these issues.

CSAGC-IDS includes two algorithms. SC-CGAN integrates CGAN with CSAM and CNN to fuse conditional information, capture feature dependencies, generate high-quality data. CSCA-CNN for traffic classification, which integrates CAM and CSL to extract deep features from complex and high-dimensional data, assign higher costs to minority classes to reduce bias caused by data imbalance. Finally, enhancing interpretability of the model provided explanations for the decision-making processes.

Based on this paper, there are several prospective directions for future work. Firstly, enhancing robustness~\cite{SETHI2020,Merzouk2022}. Secondly, reducing parameter and computational complexity~\cite{Hinton2015,Yim2017}. Thirdly, considering temporal characteristics of network traffic~\cite{WEI2023}.


\begin{credits}
\subsubsection{\ackname} I would like to express my sincere gratitude to my girl friend, Gui Tian.

\end{credits}
%
%
%
%






\end{document}